\documentclass[
    ,final            % use final for the camera ready runs
  ]
  {aipproc}

\layoutstyle{6x9}

\newcommand{\pT}        {$p_{\rm{T}}$}

\newcommand{\Deg} {$^{\rm{o}}$}

\begin{document}

\title{Rapidity Dependence of High-$p_{T}$ Suppression}

\classification{25.75.Dw,13.85Hd.}
\keywords      {Nuclear Modification Factor, gluon saturation}

\author{F.Videb{\ae}k for the BRAHMS Collaboration}{
  address={Physics Department, Brookhaven National Laboratory}
}

\begin{abstract}
The rapidity dependence of nuclear modifications factor in d-Au collisions at $\sqrt{s_{NN}}$ = 200 GeV
at RHIC is discussed.
\end{abstract}

\maketitle

%%%%%%%%%%%%%%%%%%%%%%%%%%%%%%%%%%%%%%%%%%%%
%% MAINMATTER
%%%%%%%%%%%%%%%%%%%%%%%%%%%%%%%%%%%%%%%%%%%%

\section{Introduction}
Particle production at forward rapidities in p(d)-A reactions probes partons in the target at low-x values.
At sufficient high energy, large rapidities or large nucleii the initial gluon distribution will saturate, 
and is expected to modify the particle pseudo rapidity densities, as well as the \pT~ spectra of hadrons. 
A theory based on QCD has been developed for dense low-x systems, 
termed the Color Glass Condensate (CGC) \cite{MLV94}.
This description has inspired much theoretical and experimental work,
and was also a motivation for the BRAHMS measurements
of charged hadrons at forward rapidities in 200 GeV d-Au collisions at RHIC.

\section{Results}

The data reported here were all obtained with the BRAHMS spectrometers.
The BRAHMS forward spectrometer consists of 4 dipole magnets, 5 tracking chambers, 
two Time-Of-Flight systems and a Ring Imaging Chrenkov Detector (RICH) for particle identification. 
The angular coverage of the spectrometer extends  from 2.3\Deg~ to 15\Deg~ with solid angle of  0.8 msr. 
The mid-rapidity spectrometer covers angles from 40\Deg~ and 90\Deg~. 
%%with pion identification up to \pT~ of 2 GeV/$c$.
Details of experimental setup can be found in \cite{BRAHMS_NIM}.
The collision vertex is determined from timing measurements done with a  set of symmetricaly placed
scintillator counters  around the beam pipe at 1.5, 4.15 and 6.7 meters \cite{Arsene:2004ux}.
The resolution of the vertex determination is $\approx 10$ cm. 
This set of counters also provides the minimum bias normalization. 
It is estimated that for pp collisions they record $70\%$ of the inelastic cross section.
Additional details of the setup as well as the analysis can be found in \cite{Arsene:2004ux,brahms-da_dndeta} 
where most of the data discussed here were first published.
Spectra of charged hadrons in d-Au and pp collisions at 200 GeV are presented in Fig. 1 for several pseudorapidities.
The data at $\eta=0$ and 1 are for the average of the positive and negatives charges, 
while the high rapidity data are for negative only.
Both the pp and dAu cross sections becomes steeper with increasing $\eta$.

\begin{figure}[h]
  \includegraphics[height=.3\textheight]{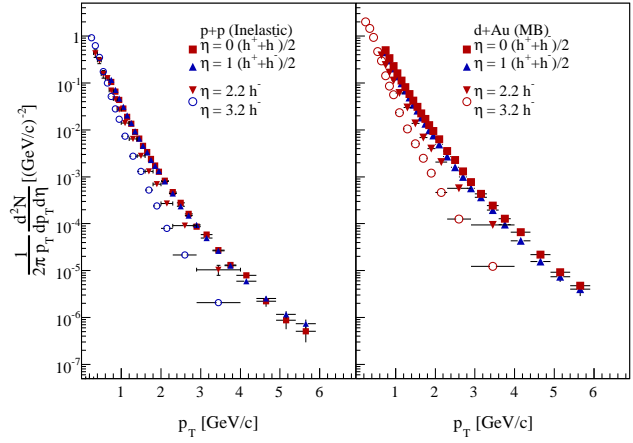}
  \caption{Spectra for dAu and pp}

\end{figure}

\begin{figure}[hb]
  \includegraphics[height=.3\textheight]{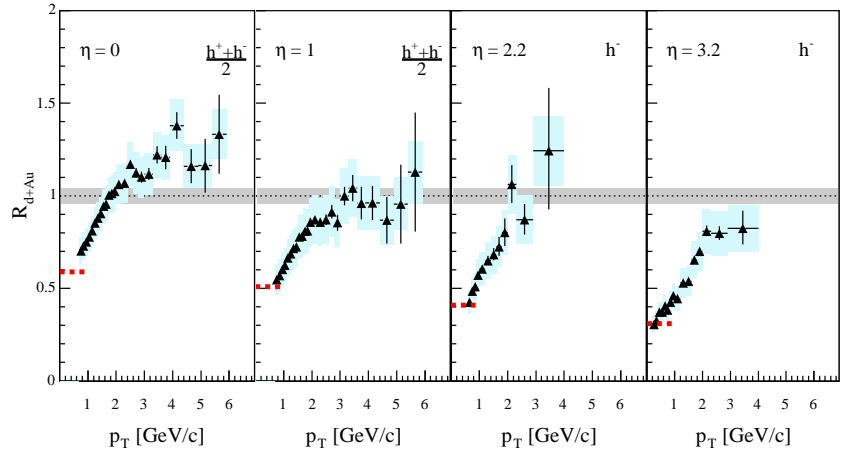}

\caption{\label{fig:ratio} Nuclear modification factor for charged
  hadrons at pseudorapidities $\eta=0,1.0,2.2,3.2$. Statistical
  errors are shown with error bars. Systematic
  errors are shown with shaded boxes 
  with widths set by the bin sizes.                                          
  The   shaded band around
  unity indicates the estimated error on the normalization to $\langle N_{coll} \rangle$. 
  Dashed lines at $p_T<1$ GeV/c show the normalized charged particle 
  density ratio $\frac{1}{\langle
  N_{coll}\rangle}\frac{dN/d\eta(d+Au)}{dN/d\eta(pp)}$.}
\end{figure}

The data collected from d-Au collisions is compared to p-p using the nuclear 
modification factor defined as: $R_{dAu}=\frac{1}{N_{coll}}\frac{\frac{dN^{dAu}}{dp_{T}d\eta}}{\frac{dN^{pp}}{dp_{T}d\eta}}$.
where $N_{coll}$ is the number of binary collisions estimated to be equal to $7.2 \pm 0.6$ for
minimum biased d+Au collisions.   The pt dependence of the factor is shown in Fig.\ref{fig:ratio}.
Each panel shows the ratio calculated at a different $\eta$ value. 
At mid-rapidity ($\eta = 0$), the nuclear modification factor exceeds 1 for
transverse momenta greater than 2 GeV/c in a similar, although less pronounced way 
as Cronin's p+A measurements performed  at lower energies \cite{cronin-effect}. 

A shift of one unit of rapidity is enough to make the Cronin type enhancement disappear, and as 
the measurements are done at higher rapidities, the ratio becomes consistently smaller than 1 
indicating a suppression in dAu collisions compared to scaled pp systems at the same energy. 
In all four panels, the statistical errors, shown as error bars (vertical lines), are dominant specially in the most
forward measurements. 
\begin{figure}[h]
\includegraphics[height=.3\textheight]{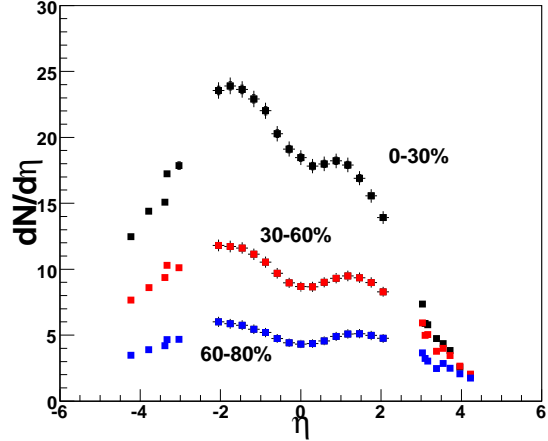}
\caption{Pseudo rapidity dependence of dN/deta in dAu for 3 centrality bins.}
\label{fig:dndeta}
\end{figure}

These results have been described within the context of the Color Glass Condensate \cite{MLV94}; 
the evolution of the nuclear modification factor with rapidity and centrality is consistent with a description of the 
Au target where  the rate of gluon fusion  
becomes  comparable with that of gluon emmission as the rapidity increases and it slows down the overall growth of the gluon
density. The measured nuclear modification factor compares the slowed down growth of the numerator to a sum of incoherent 
p+p collisions, considered as dilute systems, whose gluon densities grow faster with rapidity because of the abscence of
gluon fusion in dilute systems \cite{KKTandOthers}. Other explanations for the measured suppression
have been proposed and they also reproduce the data \cite{hwa-partonreco, Vitev, Kopeliovich}. 

Some of these other explanations  particluar focus on the observation that the 
charged particle pseudo-rapidity density distributions exhibits a change in shape vs. centrality.
This is illustrated in Fig.2 by the dashed lines at low $p_T$ where the ratios 
where obtained  from the BRAHMS data\cite{brahms-da_dndeta} 
(shown in Fig.\ref{fig:dndeta} ) and the UA5 pp results\cite{UA5}.
Thus already the overall soft spectrum shows suppression when going to forward angles.

\begin{figure}[!ht]
           {\includegraphics[height=.3\textheight]{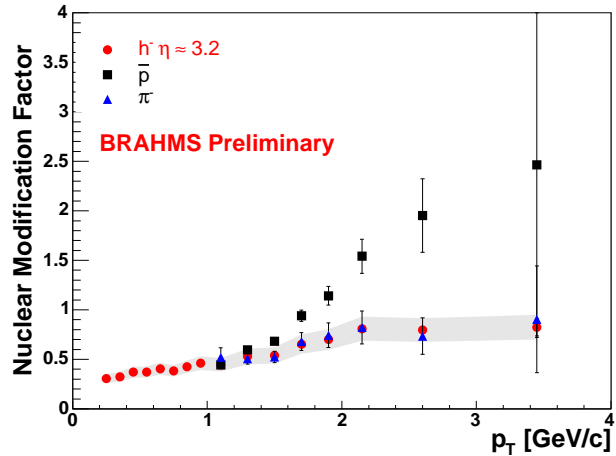}}
\caption{\label{fig:identifiedRdA}The nuclear modification factor $R_{dAu}$ calculated for
anti-protons (filled squares) and negative pions (filled triangles) at $\eta = 3.2$. The same 
ratio calcutated for negative particles  is shown with filled circles, 
and the systematic error for that
measurement is shown as grey band. }
\end{figure}

The nuclear modification factor of baryons is different from the one calculated with mesons, whenever the factor shows
the so called Cronin enhancement, baryons show a stronger enhancement. 
Such difference, seen at lower energies, has also been found at RHIC energies at all rapidities, 
in particular,  Fig. \ref{fig:identifiedRdA} presents the minimum bias 
nuclear modification $R_{dAu}$ for anti-protons and negative pions at $\eta = 3.2$. These ratios were obtained making use of ratios of raw counts of identified particles compared to
those of charged particles in each $p_{T}$ bin. 
This nuclear modification factor was calculated from the measured ratios between anti-protons 
and pions to changed particles for dAu and pp and applying
these factors to the minimum bias nuclear modification factor for negatively charged hadrons\cite{rdindia}.
%%\[ R^{\bar{p}}_{dAu} = 
%%R^{h^-}_{dAu} \frac{(\frac{\bar{p}}{h^-})^{dAu}}{(\frac{\bar{p}}{h^-})^{pp} } = 
%%\frac{1}{N_{coll}}\frac{\left.\frac{dn^{dAu}}{dp_{T}d\eta}\right)^{h^{-}}}{\left.\frac{dn^{pp}}{dp_{T}d\eta}\right)^{h^{-}}} \frac{\frac{\left.\frac{dn^{dAu}}{dp_{T}d\eta}\right)^{\bar{p}}}{\left.\frac{dn^{dAu}}{dp_{T}d\eta}\right)^{h^{-}}}}
%%{\frac{\left.\frac{dn^{pp}}{dp_{T}d\eta}\right)^{\bar{p}}}{\left.\frac{dn^{pp}}{dp_{T}d\eta}\right)^{h^{-}}}} = 
%%\frac{1}{N_{coll}}\frac{\left.\frac{dn^{dAu}}{dp_{T}d\eta}\right)^{\bar{p}}}{\left.\frac{dn^{pp}}{dp_{T}d\eta}\right)^{\bar{p}}} \]
No attempt was made to estimate the contributions from anti-lambda feed down 
to the anti-proton result. The remarkable difference between  baryons and mesons has  been related to parton recombination
\cite{hwa-partonreco} for heavy ion reactions. 
It is though surprising that this effect in the dA system at forward rapidities where only a small
soft parton component should be able to account for this increase.

In summary, particle production  from dAu and pp collisions at $\sqrt{s_{NN}}= $ 200 GeV and 
at different rapidities with the BRAHMS setup offers a window to the small-x 
components of the Au wave function. The suppression found in the particle production at high rapidities from d+Au 
collisions may be the first indication of the onset of saturation in the gluon distribution function of the Au target.

This work is supported by the Division of Nuclear Physics of the Office of Science of 
the U.S. Department of energy under contract DE-AC02-98-CH10886,
 the Danish Natural Science Research Council,
the  Research Council of Norway, the Jagiellonian University Grants 
and the Romanian Ministry of Education and Research.
I thank R.Debbe for providing valuable input to this talk and conbtribution.


\begin{thebibliography}{9}


\bibitem{MLV94} L.McLerran and R.Venugopalan,\emph{Phys. Rev.} \textbf{D49} 1994, 2233.
\bibitem{BRAHMS_NIM}J.Adamczyk et.al. \emph{Nucl. Instr. Meth.} \textbf{A499} 2003, 437 
%\cite{BRAHMS-rda}
%\cite{Arsene:2004ux}
\bibitem{Arsene:2004ux}
  I.~Arsene {\it et al.}  [BRAHMS Collaboration],
  %``On the evolution of the nuclear modification factors with rapidity and
  %centrality in d + Au collisions at s(NN)**(1/2) = 200-GeV,''
  \emph{Phys.\ Rev.\ Lett.}  {\textbf 93}, 2004, 242303 
 %% [arXiv:nucl-ex/0403005].
  %%CITATION = NUCL-EX 0403005;%%
\bibitem{brahms-da_dndeta}
I.~Arsene at.al. \emph{Phys. Rev. Letter}, \textbf{94} 2005, 032301
\bibitem{cronin-effect} D.~Antreasyan et. al. \emph{Phys. Rev.} \textbf{D19} 1979, 764


\bibitem{KKTandOthers} D. Kharzeev, Y. V. Kovchegov and K. Tuchin \emph{Phys. Rev.} \textbf{D68},2003,  094013 ; hep-ph/0307037; 
\bibitem{UA5} G.J.Almer et.al \emph{Z.Phys.}\textbf{C33} 1986, 1

\bibitem{rdindia} R.~Debbe. proceedings CINPP05; nucl-ex/0506022

\bibitem{hwa-partonreco} R.~Hwa,C.~.B~.Yang and R.~.J.~Fries 
\emph{Phys. Rev.} \textbf{C71} 2004, 024902

\bibitem{Vitev}Jian-Wei Qiu and Ivan Vitev hep-ph/0410218; Ivan Vitev hep-ph/0506039.

\bibitem{Kopeliovich} B.Z. Kopeliovich {\it et al.} hep-ph/0501260.




\end{thebibliography}
\end{document}